# Understanding the Twitter Usage of Humanities and Social Sciences Academic Journals

**Aravind Sesagiri Raamkumar**
*Nanyang Technological University, Singapore.*
aravind002 @ntu.edu.sg

**Mojisola Erdt**
*Nanyang Technological University, Singapore.*
mojisola.erdt @ntu.edu.sg

**Harsha Vijayakumar**
*Nanyang Technological University, Singapore.*
hvijayakumar @ntu.edu.sg

**Edie Rasmussen**
*University of British Columbia, Canada.*
edie.rasmussen@ubc.ca

**Yin-Leng Theng**
*Nanyang Technological University, Singapore.*
tyltheng @ntu.edu.sg

## ABSTRACT
Scholarly communication has the scope to transcend the limitations of the physical world through social media's extended coverage and shortened information paths. Accordingly, publishers have created profiles for their journals in Twitter to promote their publications and to initiate discussions with public. This paper investigates the Twitter presence of humanities and social sciences (HSS) journal titles obtained from mainstream citation indices, by analysing the interaction and communication patterns. This study utilizes webometric data collection, descriptive analysis, and social network analysis. Findings indicate that the presence of HSS journals in Twitter across disciplines is not yet substantial. Sharing of general websites appears to be the key activity performed by HSS journals in Twitter. Among them, web content from news portals and magazines are highly disseminated. Sharing of research articles and retweeting was not majorly observed. Inter-journal communication is apparent within the same citation index, but it is very minimal with journals from the other index. However, there seems to be an effort to broaden communication beyond the research community, reaching out to connect with the public.

## INTRODUCTION

Twitter can be considered as one of the contemporary and popular online social networks. As a micro-blogging system, it is relevant in both private and public communication spheres. Twitter is used for purposes such as updating current status, initiating conversations, endorsing tweet content, promoting products and even for spamming (Benevenuto, Magno, Rodrigues, & Almeida, 2010). The success of a social media platform lies in its ability to attract people from different domains and geographic locations Academicians and researchers from the scientific community are also interested in social media due to its various benefits (Sugimoto, Work, Larivière, & Haustein, 2016). Scholars have cited information dissemination as a major benefit of using Twitter (Letierce, Passant, Breslin, & Decker, 2010). It has been observed that tweets can help in predicting citations to a certain extent (Eysenbach, 2011). However, other studies advise analysts to exercise caution when using simplistic metrics such as the tweet count to evaluate research outputs (Robinson-Garcia, Costas, Isett, Melkers, & Hicks, 2017).

Social media has become a valuable marketing tool for publishers to promote research articles (Thelwall, Haustein, Larivière, & Sugimoto, 2013). Publication venues such as journals and conferences use Twitter for connecting with researchers. Studies have been conducted to investigate the usage of Twitter by researchers (Holmberg & Thelwall, 2014), conferences (Lee, Yoon, Smith, Park, & Park, 2017; Parra et al., 2016) and also journals from specific disciplines (Cosco, 2015; Kelly et al., 2016; Nason et al., 2015). It has been observed that journals with social media profiles (for e.g., Twitter) had higher academic metrics (Wong, Piraquive, & Levi, 2018). However, there has been a distinct lack of studies focusing on the usage of Twitter by journals from a broader outlook i.e., journals from different disciplines.

This study investigates the Twitter usage of humanities and social sciences (HSS) journals. In this study, we have concentrated on the soft sciences with a plan to compare the findings with the traditional science and engineering disciplines in future studies. The required data for this study was extracted from the Master Journal List (MJL)[1] and Twitter. Since the overall objective is to understand the usage dynamics of HSS journals in Twitter from a broader perspective, the following exploratory research questions are investigated in the study.

---

[1] Master Journal List of Clarivate Analytics http://ip-science.thomsonreuters.com/mjl/

- **RQ1**: What types of actions on Twitter are prevalent among HSS journals?
- **RQ2**: What is the nature of Twitter conversations among the HSS journals?
- **RQ3**: What is the prevalent network structure in Twitter communication graphs of the HSS journals?
- **RQ4**: Which Twitter accounts act as hubs and authorities in the communication graphs of the HSS journals?

**RELATED WORK**

Several studies have investigated the social media usage of journals. Twitter is an important platform for journals as studies have shown that journals with Twitter accounts have higher number of tweets and citations of their articles, when compared with other journals (Ortega, 2017). Other studies have focused on the Twitter usage of journals from specific disciplines. Findings from these studies show that the amount of Twitter accounts varies a lot across different research areas. About 14% of dermatology journals (Amir et al., 2014), 25% of urological journals (Nason et al., 2015), 28% of radiology journals (Kelly et al., 2016), and 28% of medical journals (Cosco, 2015) were found to have Twitter accounts, compared to 44% of medical journals (Kamel Boulos & Anderson, 2014). Also, 17% of the top 10 2010 JIF journals from multiple disciplines (Kortelainen & Katvala, 2012), were found to have Twitter accounts. Some studies have focussed on the correlation of Twitter and bibliometrics indicators. Even though there was no strong relation with the Journal Impact Factor (JIF) for urological journals (Nason et al., 2015), the authors found the Twitter presence of journals to be positively associated with the JIF in radiology journals (Kelly et al., 2016). For medical journals, it was found that journals with high JIF had a greater number of Twitter followers than others (Cosco, 2015). Besides maintaining accounts, which are often used to share news and articles, some journals have started to ask authors to create Wikipedia articles (Butler, 2008; Maskalyk, 2014), and to provide so-called tweetable abstracts that journals can use to promote papers (Darling, Shiffman, Côté, & Drew, 2013). Some journals have created online journal clubs to encourage community discussions about articles (Chan et al., 2015). Most use Twitter (Chan et al., 2015; Gardhouse, Budd, Yang, & Wong, 2017; Leung, Siassakos, & Khan, 2015; Mehta & Flickinger, 2014; Thangasamy et al., 2014), but some use other social media tools such as blogs, live videos, and podcasts (Rezaie, Swaminathan, Chan, Shaikh, & Lin, 2015; Thoma, Rolston, & Lin, 2014).

Apart from journals, studies have been conducted on the Twitter usage of users at academic conferences as the main focus area. With more user participation in a Twitter network, the network structure might show a tendency to change. However, a 'tight crowd' network was consistently present in the case of the Association of Internet Researchers (AoIR) conference over the years (Lee et al., 2017). Also, the authoritative Twitter accounts did not change much even though there were changes in the conference theme every year. In a much larger study conducted with tweets from 16 computer science conferences (Parra et al., 2016), authors found that users were increasingly engaged in retweeting and sharing of web links over time. There was a drop in conversational tweets between users through the years, thereby changing the network structure to more of a broadcasting network. From a study of the tweeting behaviour of participants of the 2013 International Congress for Conservation Biology (ICCB) (Bombaci et al., 2016), it was found that the intended audiences, such as policy makers, government and non-government organizations, were rarely reached by tweets. Thus, it is questionable if tweeting at conferences is an effective strategy for promoting research to a wider audience other than the conference's research community.

Based on our literature review and to the best of our knowledge, there have been no studies that report about the Twitter usage and communication network of HSS journals across multiple disciplines. The current study attempts to address this gap.

**METHODOLOGY**

*Raw Data Extraction*

The Master Journal List (MJL) was selected as the main source to obtain journal titles for this study. Specifically, journal titles were extracted from the two major HSS indices – Arts & Humanities Citation (AHCI) and Social Science Citation Index (SSCI). The study was conducted on a corpus of 321,094 tweets extracted from 4,999 journal accounts of the two citation indices. Data collection details are provided below. MJL is generally acknowledged as one of the main sources for journal titles. Journals are included in MJL after a quality check process, thereby ensuring compliance to scientific standards. Hence, MJL can be considered as an authoritative source for journal titles across disciplines. Humanities and social sciences journals have been traditionally indexed under two indices in MJL, namely Arts & Humanities Citation (AHCI) and Social Science Citation Index (SSCI). All journal titles from the two indices AHCI and SSCI were manually downloaded. In total, 4,999 relevant journals were retrieved, including 1,769 journals from AHCI and 3,230 journals from SSCI. To examine whether the extracted journals have their own Twitter accounts, we searched for their titles using the search option in Twitter. After collecting the Twitter accounts of the journals in August 2016, we extracted the tweets from the Twitter accounts, using the Twitter API. A maximum of 3,000 tweets were extracted for each Twitter account due to the restrictions in the basic Twitter API service. A total of 321,094 tweets were extracted for the 4,999 journals.

## *Preparation of Data for Study*

### Tweets with URLs

The tweets which contained links to research articles were identified using a two-step process. First, the tweets containing Uniform Resource Locators (URLs) were filtered. From this filtered list, further filtering was carried out based on the presence of any one of the keywords 'doi', 'article' and 'issue' to identify the candidate tweets.

### Twitter Mentions and Conversations

In Twitter terminology, *mention* is an instance of tagging/mentioning another Twitter user in a tweet. For example, if *user A* wants to start a discussion with *user B*, the "@" is used to tag *user B* in the tweet. It should also be noted that when a user retweets the tweet of another user, Twitter automatically adds the characters "RT @user_account" in the tweet. Hence, mentions are naturally present in retweets. From the full extract of tweets, only the tweets containing mentions were first filtered. From the filtered tweets, the Twitter account name (Twitter handle) and the mentions data were extracted. A combination of Twitter account name and mention is usually referred to as a *conversation*. A single tweet could contain multiple mentions.

### Communication Graphs

The network analysis tool Gephi was used for addressing RQ3 and RQ4. The community detection algorithm ForeAtlas2 (Jacomy, Venturini, Heymann, & Bastian, 2014) was used in Gephi to align the communities in the graphs. The nodes in the graphs were sized based on their betweenness centrality (Freeman, 1977) values. Betweenness centrality is a measure of the centrality of nodes in a graph. Communication graphs were generated with the conversations data. In these graphs, the source node is the Twitter user account while the target node is the mention. This type of graph is referred to as a *directed graph* since the direction of communication is from the source to the target. Two separate directed graphs were generated using the Gephi tool. After the data was loaded in Gephi, we filtered out the nodes which had a degree of 1 to avoid sparsity in the graphs.

## RESULTS

### *Presence and Activities of HSS Journals in Twitter*

In Table 1, statistics related to the journals presence in Twitter are listed. Specifically, counts and percentages of journals and their tweets, retweets, mentions and presence of both general URLs & research article URLs in tweets. The percentage of journals with Twitter accounts is around 7% to 9% across the two indices with AHCI having the higher percentage of journals with Twitter accounts (8.99%). The low percentages indicate that a clear majority of HSS journals are yet to establish a presence in Twitter. Secondly, the presence is also affected by publication houses which are based out of countries where Twitter is banned. Retweeting is considered as one of the major activities in Twitter (Yang et al., 2010). Journals from SSCI had the highest number of retweets (23.39% of total) closely followed by AHCI (22.01% of total). Apart from retweeting, URL sharing is also one of the major activities of Twitter users (Java, Song, Finin, & Tseng, 2007). SSCI journals had a higher number of tweets with URLs (89.89% of total), followed by AHCI (86.76% of total). The presence of URLs in a tweet indicates that the journal is posting the URL of a research paper or website (e.g. news articles, blogs). The percentage of tweets containing links to research articles is not substantial (less than 15% of total), with SSCI (11.03% of total) having a better percentage than AHCI (4.98%).

| Entity | AHCI (*n*) | SSCI (*n*) |
|---|---|---|
| Journals [A] | 1769 | 3230 |
| Journals with Twitter Accounts [B] | 159 (8.99% of A) | 249 (7.71% of A) |
| Extracted Tweets [C] | 145419 | 175675 |
| Journals with Public Tweets [D] | 159 (100% of B) | 249 (100% of B) |
| Retweets (RTs) [E] | 32002 (22.01% of C) | 41096 (23.39% of C) |
| Tweets containing URLs [F] | 126172 (86.76% of C) | 157918 (89.89% of C) |
| Tweets containing Links to Articles [G] | 7240 (4.98% of C) | 19370 (11.03% of C) |

**Table 1. Journals' Twitter Presence and Activities Statistics**

*Twitter Conversations*

In Table 2, statistics related to the conversations of HSS journals in Twitter are listed for the two indices. The number of conversations was the highest for SSCI ($n=121,099$) corresponding to the higher number of tweets (refer Table 1). Since only the tweets containing mentions data were considered, some of the journals had zero conversations even though tweets could have been posted by these accounts. For instance, out of the 249 journal accounts with tweets in SSCI (refer Table 1), 242 accounts had mentions data.

| Entity | AHCI ($n$) | SSCI ($n$) |
|---|---|---|
| Total conversations [A] | 103181 | 121099 |
| Unique journal accounts initiating conversations [B] | 152 | 242 |
| Unique mentions [C] | 31040 | 32819 |
| Conversations where mentions are AHCI journals [D] | 13812 (13.39% of A) | 70 (0.06% of A) |
| Conversations where mentions are SSCI journals [E] | 63 (0.06% of A) | 16048 (13.25% of A) |
| Unique AHCI journal mentions [F] | 142 (93.42% of B) | 18 |
| Unique SSCI journal mentions [G] | 18 | 230 (95.04% of B) |
| Conversations where account and mention are same [H] | 13015 (12.61% of A) | 14398 (11.89% of A) |

**Table 2. Journals' Twitter Conversations Statistics**

In a conversation, the mention could be any Twitter account. To ascertain the level of interaction with other journals, we identified the mentions which were journal accounts. For journals within the same index, AHCI and SSCI conversations had similar percentages (13.39% for AHCI, 13.25% for SSCI). This finding indicates that journals do not seem to be interacting with other journals in Twitter to any great degree. In fact, interactions with journals from other index is very low with percentages below 1%.

However, most of the journals within the indices are involved in the conversations. For instance, 95.04% of all SSCI journals and 93.42% of all AHCI journals are presented as mentions in the conversations respectively. To investigate further, we identified the conversations where the user account and mentions were the same. We found that these types of conversations were around the same range for both indices (11.89% for SSCI and 12.61% for AHICI of the total conversations respectively). This finding is quite understandable since there are not many benefits for journals to tag their own Twitter account in their tweets

*Communication Graphs*

Statistics of the two communication graphs generated for AHCI and SSCI journals Twitter conversations are listed in Table 3. The node count in Table 3 corresponds to the number of conversations for each index in Table 2. SSCI ($n=4,656$) and AHCI ($n=4,280$) had similar number of nodes. In the case of average degree of a node, AHCI ($a=2.84$) and SSCI ($a=3.08$) were relatively close to each other. This finding indicates that even though the number of nodes is lower in the graph, each node has more connections with other nodes in the graph. Modularity ($m$) in a graph is an indication of how well a graph decomposes into modular communities. A high modularity hints at a complex internal structure with multiple sub-graphs (communities) densely connected to each other in the graph. The modularity value should be analysed together with the number of resultant communities ($c$) and total number of nodes in the graph ($n$). The modularity value is almost the same for the two graphs ($m=0.53$ for AHCI and $m=0.52$ for SSCI). Resultantly, the number of communities is also comparable for AHCI ($c=11$) and SSCI ($c=13$).

| Metric | AHCI | SSCI |
|---|---|---|
| Number of Nodes ($n$) | 4,280 | 4,656 |
| Number of Edges ($e$) | 12,194 | 14,353 |
| Average Degree ($a$) | 2.84 | 3.08 |
| Modularity ($m$) | 0.53 | 0.52 |
| Number of Communities ($c$) | 11 | 13 |

**Table 3. Journals' Communication Graphs Statistics**

The two graphs generated for the citation indices are illustrated in Figures 1 and 2. The node colour indicates the community to which they belong. The nodes size has been adjusted based on the betweenness centrality values of the nodes. We have added representative labels for the different communities based on the topic commonality in the profile descriptions provided in the Twitter accounts (journals). As an aid to interpret the graphs, we identified the top 20 nodes with the highest in-degrees and out-degrees respectively in Tables 4 and 5. Out-degree of a node is the number of edges where the node is the source node while in-degree is the number of edges where the node is a target node. Naturally, the nodes with the highest out-degrees will be the journal Twitter accounts while the nodes with the highest in-degrees could be of any type. Accordingly, we have listed the journal titles for the nodes with the highest out-degrees while we have identified type of the nodes with highest in-degrees. Even though out-degree is an indication of the journal's outreach in Twitter, it is important to recognize the popularity of the journals within the traditional citation network. This can be achieved by comparing the degree values with the journal impact factor (JIF). We have not listed the JIF values in these tables since interpreting JIF with different disciplines is not meaningful. However, we have listed the quartiles data of the journals based on the JIF. For example, if the journal belongs to the first quartile, it is considered to be among the top 25 percentile of journals in that discipline. Therefore, the prestige of journals can be ascertained with the JIF quartiles data. For AHCI journals, JIF data is not provided in the MJL, hence JIF quartiles column is not applicable for Table 4.

AHCI Graph

In Figure 1, the AHCI graph is represented by multiple topics including philosophy, history, arts, architecture, film and literature. Most of the Twitter accounts in this graph are related to the literature topic. Also, the communities adjacent to each other seem to be related - (1) architecture & art and (2) philosophy & history. This graph can be classified as a *community clusters* graph since there are multiple communities with minimal interspersed nodes. From Table 4, we observe that magazines and news portals are most frequently referenced as mentions in AHCI Twitter conversations. The top two nodes are the news portals Nytimes and Guardian, while six other nodes are of the magazine type in the top 20. However, these magazines are all within the scope of AHCI topics unlike the news portals which cover general topics. Libraries (@nypl, @ElectricLit) and museums (@metmuseum, @MusuemModernArt) are also present, thereby indicating substantial interactions with accounts outside academia. Paris Review is the only journal with a high in-degree value in this list and this journal's node can be clearly identified as the biggest node in Figure 1 due to its high betweeness centrality. The top 20 in-degree nodes indicate that AHCI journals tend to interact most with news and magazine-related Twitter accounts. The top 20 out-degrees nodes comprise of mostly literature and arts journals.

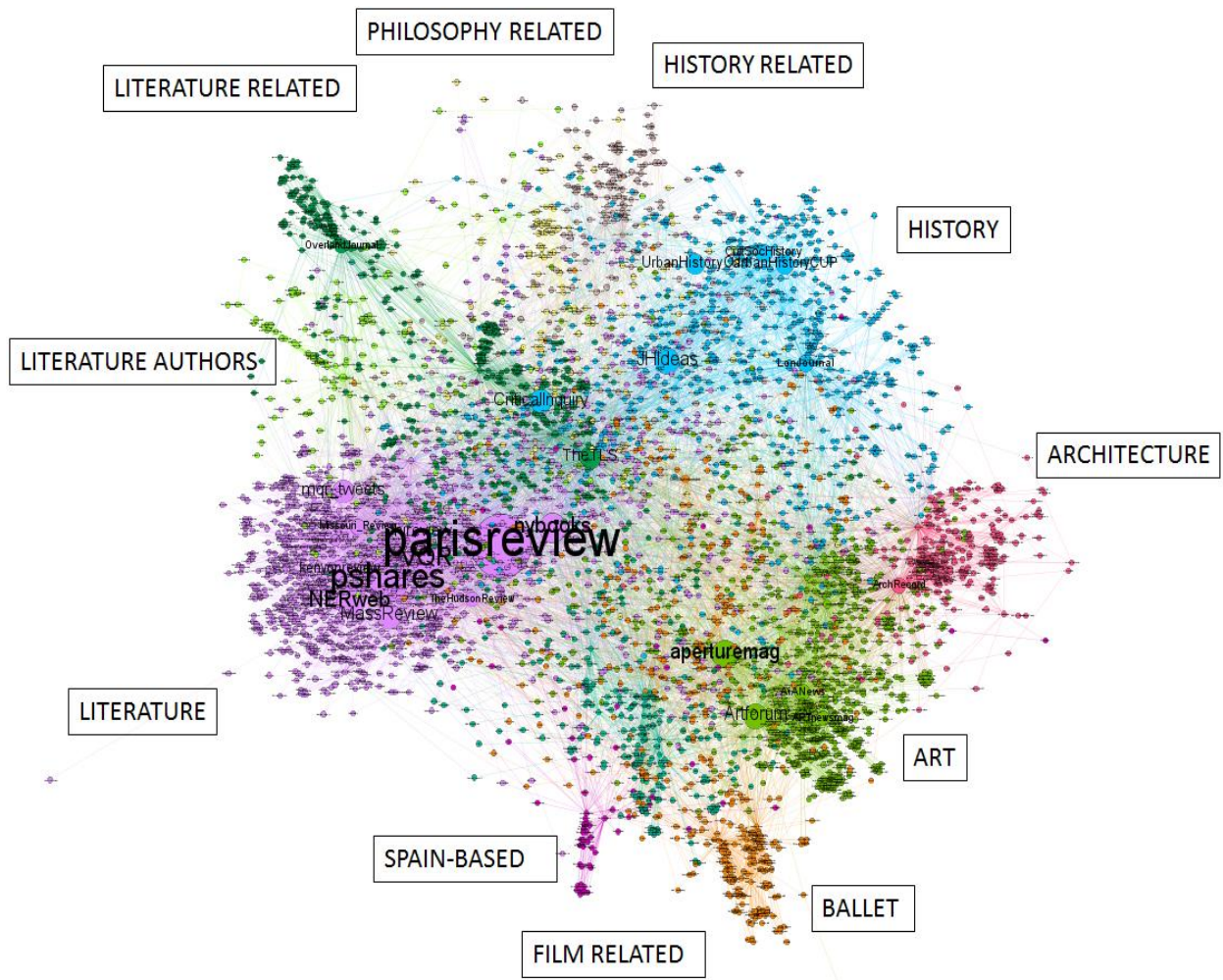

**Figure 1. AHCI Graph**

| Top 20 Nodes with Highest In-degree | | | Top 20 Nodes with Highest Out-degree | | |
|---|---|---|---|---|---|
| Twitter Account | Account Type | $in^*$ | Twitter Account | Journal Full Name | $out^+$ |
| Nytimes | News | 39 | VQR | Virginia Quarterly Review | 549 |
| Guardian | News | 30 | AiANews | Art In America | 468 |
| NewYorker | Magazine | 29 | Missouri_Review | Missouri Review | 439 |
| Parisreview | Journal | 23 | Wasafiri1 | Wasafiri | 422 |
| Nypl | Library | 22 | Artforum | Artforum International | 391 |
| Sharethis | Social Bookmarking | 22 | kenyonreview | Kenyon Review | 366 |
| LAReviewofBooks | Magazine | 21 | TheHudsonReview | Hudson Review | 356 |
| Metmuseum | Museum | 20 | ArchRecord | Architectural Record | 346 |
| Tate | Art Gallery | 19 | NERweb | New England Review-Middlebury Series | 269 |
| LRB | Magazine | 18 | LonJournal | London Journal | 263 |
| PoetryFound | Magazine | 18 | BurlingtonMag | Burlington Magazine | 258 |
| nybooks | Magazine | 18 | AD_books | Architectural Design | 239 |
| TheAtlantic | Magazine | 18 | nybooks | New York Review Of Books | 232 |
| ElectricLit | Digital Library | 18 | TheTLS | Tls-The Times Literary Supplement | 231 |
| MuseumModernArt | Museum | 16 | TheAmScho | American Scholar | 225 |
| OUPAcademic | Publisher | 16 | JCLJournal | Journal Of Commonwealth Literature | 225 |
| YouTube | Video Sharing | 16 | Apollo_magazine | Apollo-The International Art Magazine | 205 |
| gmailcom | Email | 16 | NOReview | New Orleans Review | 195 |
| PublishersWkly | Magazine | 15 | MassReview | Massachusetts Review | 192 |
| thelithub | Not Active | 15 | ARTnewsmag | Artnews | 190 |

*Note: $in^*$ refers to in-degree value; $out^+$ refers to out-degree value*

**Table 4. Top 20 In-degree and Out-degree Nodes in AHCI Graph**

SSCI Graph

The SSCI graph, which has the highest number of communities ($c$=13) among the two graphs (refer Table 3), is illustrated in Figure 2. Interestingly, most of the communities are of different topics although the adjacent communities are related, similar to AHCI. The community related to feminism/women's studies has the largest nodes (@FeministReview, @AFSJournal) and this community is spread across other communities related to psychology and education. Similar to AHCI, this graph can also be classified as a *community clusters* graph since there are multiple communities and minimal interspersed nodes even though the feminism community is an exception.

Among the top 20 nodes with highest in-degrees, news portals (e.g., @nytimes, @washington) and magazines (e.g., @TheEconomist, @TheAtlantic) have a significant presence as already seen with AHCI. Unlike AHCI, news portals' ($n$=7) presence is more than magazines ($n$=4). Also, the magazines are mainstream magazines and not from academic publishers. Only two journals (@TheSocReview, @SAGEsociology) are present in this top 20 in-degree nodes list. The journals in the top 20 nodes with highest out-degrees are spread across different disciplines. From the JIF quartiles data, it can be ascertained that only three journals (@AmJNurs, @ AmEthno, @ASQJournal) are from the first quartile. There are more second quartile ($n$=7), third quartile ($n$=6) and fourth quartile ($n$=4) journals in this list. Therefore, the outreach is performed by journals from all tiers. Also, there is no overlap among the journals in the in-degree and out-degree list.

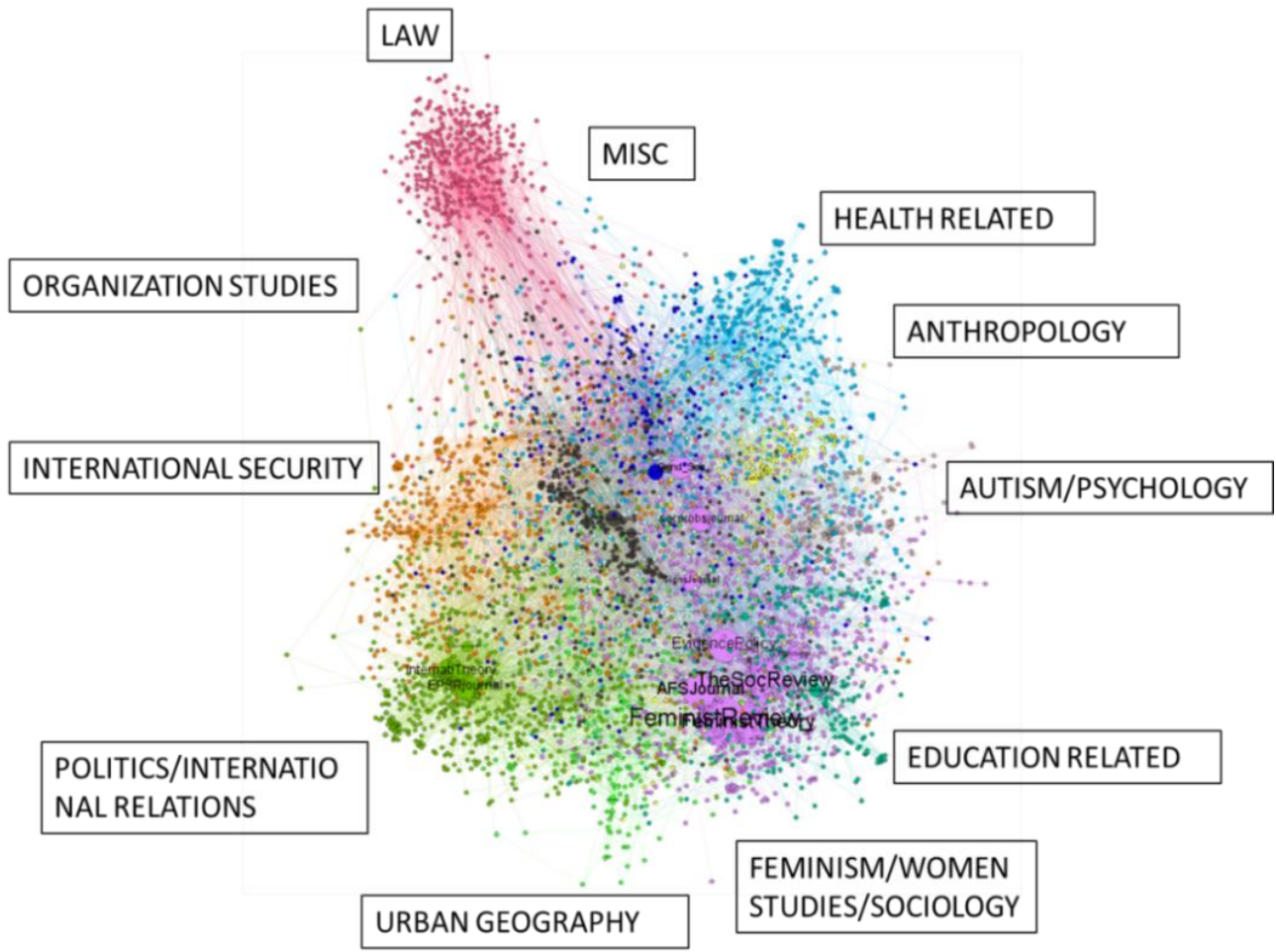

**Figure 2. SSCI Graph**

| Top 20 Nodes with Highest In-degree | | | Top 20 Nodes with Highest Out-degree | | | |
|---|---|---|---|---|---|---|
| Twitter Account | Account Type | in* | Twitter Account | Corresponding Journal Title | out+ | JQ^ |
| nytimes | News | 67 | InternatlTheory | International Theory | 338 | 3 |
| guardian | News | 60 | AFSJournal | Australian Feminist Studies | 336 | 3 |
| washingtonpost | News | 46 | AmJNurs | American Journal Of Nursing | 277 | 1 |
| TheEconomist | Magazine | 41 | EPSRjournal | European Political Science Review | 268 | 2 |
| TheAtlantic | Magazine | 37 | po_qu | Political Quarterly | 260 | 4 |
| WSJ | News | 34 | EvidencePolicy | Evidence & Policy | 255 | 2 |
| raulpacheco | Researcher | 27 | BulletinAtomic | Bulletin Of The Atomic Scientists | 253 | 4 |
| chronicle | News | 27 | terpolv | Terrorism And Political Violence | 251 | 2 |
| sharethis | Social Bookmarking | 25 | BASeditors | Business & Society | 242 | 2 |
| NPR | Media Organization | 23 | AmEthno | American Ethnologist | 242 | 1 |
| NewYorker | Magazine | 23 | govandopp | Government And Opposition | 215 | 3 |
| HuffingtonPost | News | 23 | ERSjournal | Ethnic And Racial Studies | 211 | 2 |
| TheSocReview | Journal | 23 | psychmag | Psychologist | 208 | 4 |
| SAGEsociology | Journal | 23 | CMPjournal | Culture Medicine And Psychiatry | 195 | 3 |
| timeshighered | Magazine | 23 | LAPerspectives | Latin American Perspectives | 186 | 3 |
| YouTube | Video Sharing | 22 | mgmt_learning | Management Learning | 185 | 2 |
| TIME | Magazine | 22 | hhrjournal | Health And Human Rights | 173 | 3 |
| Slate | Magazine | 22 | Editor_IES | Irish Educational Studies | 166 | 4 |
| wordpressdotcom | Content Management | 21 | socprobsjournal | Social Problems | 162 | 2 |
| ConversationUK | News | 21 | ASQJournal | Administrative Science Quarterly | 154 | 1 |

*Note: in\* refers to in-degree value; out+ refers to out-degree value; JQ^ refers to JIF quartile*

**Table 5. Top 20 In-degree and Out-degree Nodes in SSCI Graph**

## DISCUSSION

The presence of academic journals in Twitter is not yet substantial with AHCI having a slightly larger presence than SSCI journals. There is not much difference between the indices in terms of having a presence in Twitter. As Twitter's usage and popularity increases, more journals are expected to join the social media platform. However, the participation of journals in social media might be constrained by monetary and manpower availability factors. For the journals in this study, the retweeting frequency does not seem to exceed the general statistics reported in earlier studies (Holmberg & Thelwall, 2014) with SSCI journals having the highest percentage (23.39% of total tweets) of the two indices. In terms of URL sharing in tweets, journals exceed the sharing percentage from earlier studies (Raamkumar, Pang, & Foo, 2016). In fact, 89.89% of SSCI journals' tweets contain URLs. Even though URLs were found in most of the tweets, the percentage of tweets containing links to research articles was found to be low with AHCI being the lowest (4.98% of total tweets). The main reason could be the limited number of articles published per journal issue. For instance, most journals post a new issue every quarter and the new issue might comprise of eight to ten articles. Hence, the tweet count in such cases would obviously appear very small when compared to the total count of tweets.

In Twitter, there has been an apparent lack of interpersonal communication due to the predominance of retweeting and URL sharing by users (Parra et al., 2016). Therefore, in this context, understanding the volume of interpersonal communication is important for journals. This analysis was performed by extracting the mentions data from the tweets. After filtering out the retweets in the extract, it was found that AHCI tweets contained mentions in 22% of their total tweets. When the analysis is extended to composition of the conversations, there are new insights gained from the data (refer to Table 2). In the mentions data, the accounts which are journals were identified so that the extent of inter-journal communication could be ascertained. SSCI mentions had the highest percentage of references to SSCI journals themselves at 13.25% of the total conversations. In fact, SSCI and AHCI are closer in this aspect. This indicates that the journals from these indices are communicating more with non-academic Twitter accounts than HSS journal accounts. At a consolidated level, we found a reasonable percentage of tweets containing mentions data; however, most of the mentions were not academic HSS journal accounts. This clearly shows that inter-journal communication in Twitter is not substantial across the two indices.

AHCI and SSCI mentions graph structure largely resemble a *community clusters* network where there are multiple communities with dense intercommunication. There were a few exceptional cases such as the feminism community in SSCI which had nodes spread across a few other communities. Authoritative nodes in the mentions graphs are the Twitter accounts which are centripetal in nature i.e., most of the conversations are directed towards these nodes. After classifying the node type in the top 20 nodes with the highest in-degrees, it was evident that news portals such as the New York Times and Guardian along with magazines such as the NewYorker and The Atlantic were the most authoritative Twitter accounts in the two graphs. This finding underlines the influence of these sources on HSS academic journals. The presence of YouTube as an authoritative source was not surprising due to the abundance of educational/informational videos hosted by the service. Subject-specific magazines such as the New Scientist and The Economist were also prevalent in the top 20 in-degree nodes list of the two indices respectively. The minimal presence of journals in the list of authoritative nodes was a surprising finding. AHCI and SSCI had just a total of three journals combined. This finding not only shows that there are very few journals in the authoritative nodes list of the indices but also there seems to be a conscious effort among journals to propagate tweets of non-academic sources. The level of outreach was ascertained from the top 20 out-degrees nodes in the mentions graphs. These nodes are centrifugal in nature and hence, they are referred as hubs. In these top 20 lists, AHCI and SSCI had journals from different disciplines. On the question of ascertaining whether the top journals in Twitter were also top journals in the citation network, SSCI had journals from all four JIF quartiles in the top 20 list.

There are a few limitations in this study. By considering journals exclusively from MJL, we might be missing some other important journals which are indexed elsewhere. By the end of 2017, Twitter increased the character count in tweets to 280 from the earlier 140 characters. Hence, users can post more descriptive content and tag more users in their tweets. Therefore, this user-interface (UI) level change could possibly alter the usage dynamics of Twitter users.

## CONCLUSION

In this paper, our goal was to understand the Twitter usage dynamics of HSS journals sourced from the citation indices in MJL. Using a corpus of 321,094 tweets which were extracted for the 4,999 journal accounts, analysis was performed at activity, tweet composition and structural levels. The tweets of the journals from AHCI and SSCI were analysed separately. Results show that URL sharing was a major activity performed by academic journals. HSS journals seem to focus more on initiating conversations with other Twitter accounts, although most were of a non-academic nature. Inter-journal communication seemed to be largely restricted to the journals within the same index. Tweets from public news portals and magazines were heavily disseminated by journal accounts, thereby identifying non-academic sources as the authoritative sources in the communication graphs of the two indices. Among the journals that initiate conversations, HSS journals irrespective of their standing in the citation network, are involved in Twitter outreach. In future studies, we plan to conduct in-depth investigations on whether the activities performed by HSS journals in Twitter lead to increased citations for constituent research articles. In addition, we plan to conduct similar analysis on Science Citation Index Expanded (SCIE) journals to compare the findings with this study. We hope that this study's findings are of value to publishers, researchers and social media campaign managers, who are responsible for the marketing, branding, and promotion of a journal's services on Twitter.

**ACKNOWLEDGMENTS**

The research project "Altmetrics: Rethinking And Exploring New Ways Of Measuring Research" was supported by the National Research Foundation, Prime Minister's Office, Singapore under its Science of Research, Innovation and Enterprise programme (SRIE Award No. NRF2014-NRF-SRIE001-019).